\pdfoutput=1
\documentclass[conference,10pt]{IEEEtran}
\usepackage{cite}

\ifCLASSINFOpdf
\else
\fi
\usepackage{array}
\usepackage{url}

\usepackage{graphicx}
\usepackage{color}
\usepackage{placeins}
\usepackage{float}
\usepackage{tabularx,colortbl}

\usepackage{gensymb} 

\usepackage{prettyref}
\newrefformat{sec}{Sec.~\ref{#1}} 
\newrefformat{sub}{Sec.~\ref{#1}} 
\newrefformat{tab}{Table~\ref{#1}} 
\newrefformat{fig}{Fig.~\ref{#1}} 
\newrefformat{eqn}{(\ref{#1})} 
\newrefformat{app}{App.~\ref{#1}} 

\begin{document}
%
\title{Antenna Array Characterization via Radio Interferometry Observation of Astronomical Sources}



%
\author{\IEEEauthorblockN{T. M. Colegate\IEEEauthorrefmark{1}\IEEEauthorrefmark{2}, 
A. T. Sutinjo\IEEEauthorrefmark{2}, 
P. J. Hall\IEEEauthorrefmark{2}, 
S. K. Padhi\IEEEauthorrefmark{2}, 
R. B. Wayth\IEEEauthorrefmark{2}, \\
J. G. Bij de Vaate\IEEEauthorrefmark{3},
B. Crosse\IEEEauthorrefmark{2},
D. Emrich\IEEEauthorrefmark{2}, 
A. J. Faulkner\IEEEauthorrefmark{4}, 
N. Hurley-Walker\IEEEauthorrefmark{2}, \\  
E. de Lera Acedo\IEEEauthorrefmark{4},  
B. Juswardy\IEEEauthorrefmark{2}, 
N. Razavi-Ghods\IEEEauthorrefmark{4},
S. J. Tingay\IEEEauthorrefmark{2},
A. Williams\IEEEauthorrefmark{2}}
\IEEEauthorblockA{\IEEEauthorrefmark{2}International Centre for Radio Astronomy Research (ICRAR), Curtin University, \\
 GPO Box U1987, Perth, WA 6845, Australia\\
 E-mail: t.colegate@curtin.edu.au}
\IEEEauthorblockA{\IEEEauthorrefmark{3}ASTRON, Netherlands Institute for Radio Astronomy, 7990 AA, Dwingeloo, The Netherlands}
\IEEEauthorblockA{\IEEEauthorrefmark{4}
Cavendish Laboratory, University of Cambridge,  JJ Thompson Avenue, Cambridge, CB3 0HE UK}}


\maketitle

\begin{abstract}
We present an in-situ antenna characterization method and results for a ``low-frequency'' radio astronomy engineering prototype array, characterized over the 75--300\,MHz frequency range. The presence of multiple cosmic radio sources, particularly the dominant Galactic noise, makes in-situ characterization at these frequencies challenging; however, it will be shown that high quality measurement is possible via radio interferometry techniques. This method is well-known in the radio astronomy community but seems less so in antenna measurement and wireless communications communities, although the measurement challenges involving multiple undesired sources in the antenna field-of-view bear some similarities. We discuss this approach and our results with the expectation that this principle may find greater application in related fields.

\end{abstract}


%
\IEEEpeerreviewmaketitle

\section{Introduction}
\begin{figure}[h]
\begin{center}
\includegraphics[width=\columnwidth]{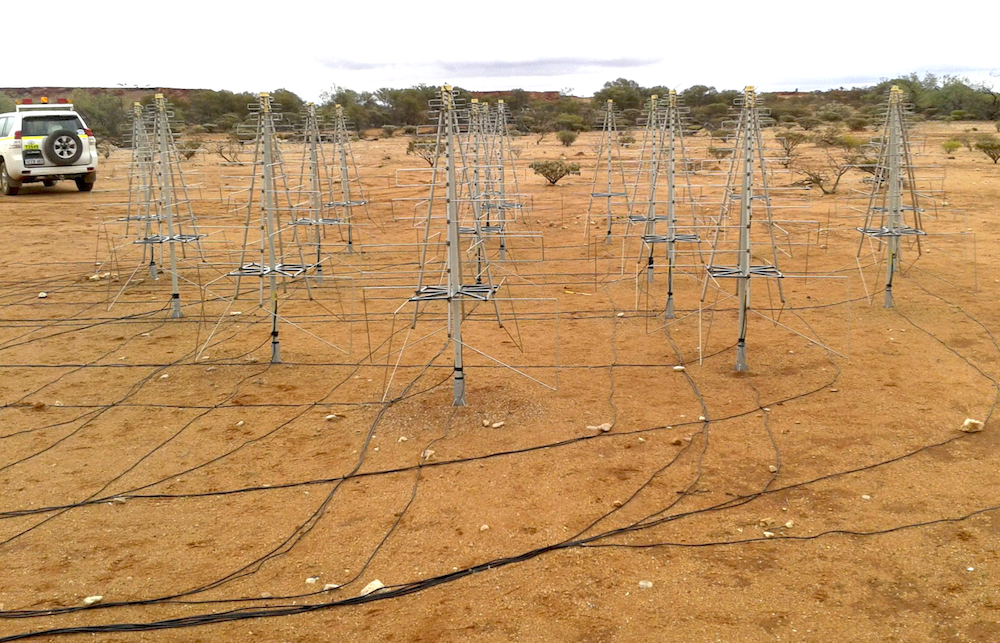}
\caption{An LFAA prototype array of 16 SKALA antennas. This system is referred to as Aperture Array Verification System 0.5 (AAVS\,0.5), and is an initiative of the SKA Array Design and Construction Consortium (AADCC). It was constructed by the International Centre for Radio Astronomy Research (ICRAR), University of Cambridge and the Netherlands Institute for Radio Astronomy (ASTRON). The SKALAs are placed in a pseudo-random configuration within an area 8\,m  in diameter.}
 \label{fig:jan2014}
\end{center}
\end{figure}

As with any antenna engineering project, characterization of low-frequency radio astronomy arrays (referred to as ``aperture arrays'' to distinguish them from dish antennas) is required to validate that the design meets requirements and that modeling tools involved in the process are reliable. Low-frequency aperture arrays (LFAA) are particularly challenging to measure for a combination of reasons: they are difficult to point mechanically, are more closely coupled to the environment (see array placement in Fig.~\ref{fig:jan2014}) and their physical area is large for a given directivity (scales by $\lambda^2$). As a consequence of the latter, larger distances (scales by the square of the array diameter) are required to satisfy the far-field condition. Without specialized facilities, measurements of LFAAs in an anechoic chamber or outdoor range are not practical.\footnote{Unmanned aerial vehicles enable outdoor measurements~\cite{VirLin14}, however the airborne height required to meet the far-field condition limits the size of the antenna or array under test (AUT).}
The capability of testing larger-diameter LFAAs will be important in the pre-construction work of the Square Kilometre Array (SKA) radio telescope~\cite{VaaLer11}.

In-situ LFAA measurements offers a number of obvious advantages: no bespoke facility is required and the array is situated in the intended environment. 
Fig.~\ref{fig:jan2014} illustrates the AUT: an array of 16 dual-polarized log-periodic ``SKALA'' antennas~\cite{EloyICEAA12} deployed at the Murchison Radio-astronomy Observatory (MRO) in the Mid West region of Western Australia~\cite{AAVS05_ICEAA13}. 
Unlike measurements in an anechoic chamber where the operator has full control of the RF sources, astronomical sources are beyond our control. This is especially true in low-frequency radio astronomy, where the spatially extended component of the emission from the plane of the Milky Way Galaxy is the dominant source of noise~\cite{Kraus_RA_1986_ch8}. 
 Meaningful LFAA measurement, therefore, requires exclusion of the diffuse Galactic emission while maintaining responsiveness to previously well-characterized, point-like extragalactic sources; this may be accomplished by cross-correlating the output voltages of pairs of AUTs which are spaced with sufficient distance such that the Galactic noise is uncorrelated. This concept will be discussed in Sec.~\ref{sec:theory}. Measurement results are and conclusions are presented in Secs.~\ref{sec:results} and \ref{ref:concl} respectively.

\section{Background Theory}
\label{sec:theory}
\subsection{Measurand}
\label{sec:meas}
The sensitivity of a radio telescope is quantified by the ratio of its effective aperture area ($A_{\rm}$ in $\rm m^2$) to the system noise temperature ($T_{\rm sys}$ in K):
\begin{equation}
A/T=\frac{A_{\rm e}}{T_{\rm sys}}.
\label{eqn:AonT-simple}
\end{equation}
This figure-of-merit is a key term in the expression for the telescope output signal-to-noise (S/N) ratio; many readers will recognize that it is similar and is convertible to $G/T$ which is common in telecommunications. The antenna beam pattern may be obtained by measuring $A/T$ along the apparent trajectory of a suitable astronomical source. This assumes that the sky noise does not change $T_{\rm sys}$ appreciably during the observation; in practice, this is a reasonable approximation.

\subsection{Measurement Method}
\label{sec:Method}
This subsection presents a simplified version of how $A/T$ is obtained in radio interferometry. Interested readers are referred to standard textbooks such as~\cite{taylor_synthesis_1999} and~\cite{tms01} for more complete discussions. Let us assume an interferometer involving 3 single-polarized antennas (or antenna arrays as the case may be) observing a point source with flux~$S$ (in W/m$^{2}$/Hz or Jy) in its phase center (in this context ``phase center'' refers to that position in the sky for which the interferometer is intended to point to by applying relative phase shifts between its elements). The measured cross-correlation products are:
\begin{eqnarray}
V_{12}=\tilde{a}_{1}\tilde{a}_{2}^{*}S \nonumber \\
V_{13}=\tilde{a}_{1}\tilde{a}_{3}^{*}S \nonumber \\
V_{23}=\tilde{a}_{2}\tilde{a}_{3}^{*}S,
\label{eqn:3ant}
\end{eqnarray}
where $\tilde{a}$ is referred to as complex ``gain'' (voltage quantity) containing a combination of both antenna and receiver electronic gains. Assuming $S$ is known, the complex ``gains'' may be solved from these 3 equations. This is typically done by solving the amplitudes and phases separately~\cite{taylor_synthesis_1999}. 

\begin{figure}[h]
 \begin{center}
 \includegraphics[width=2in]{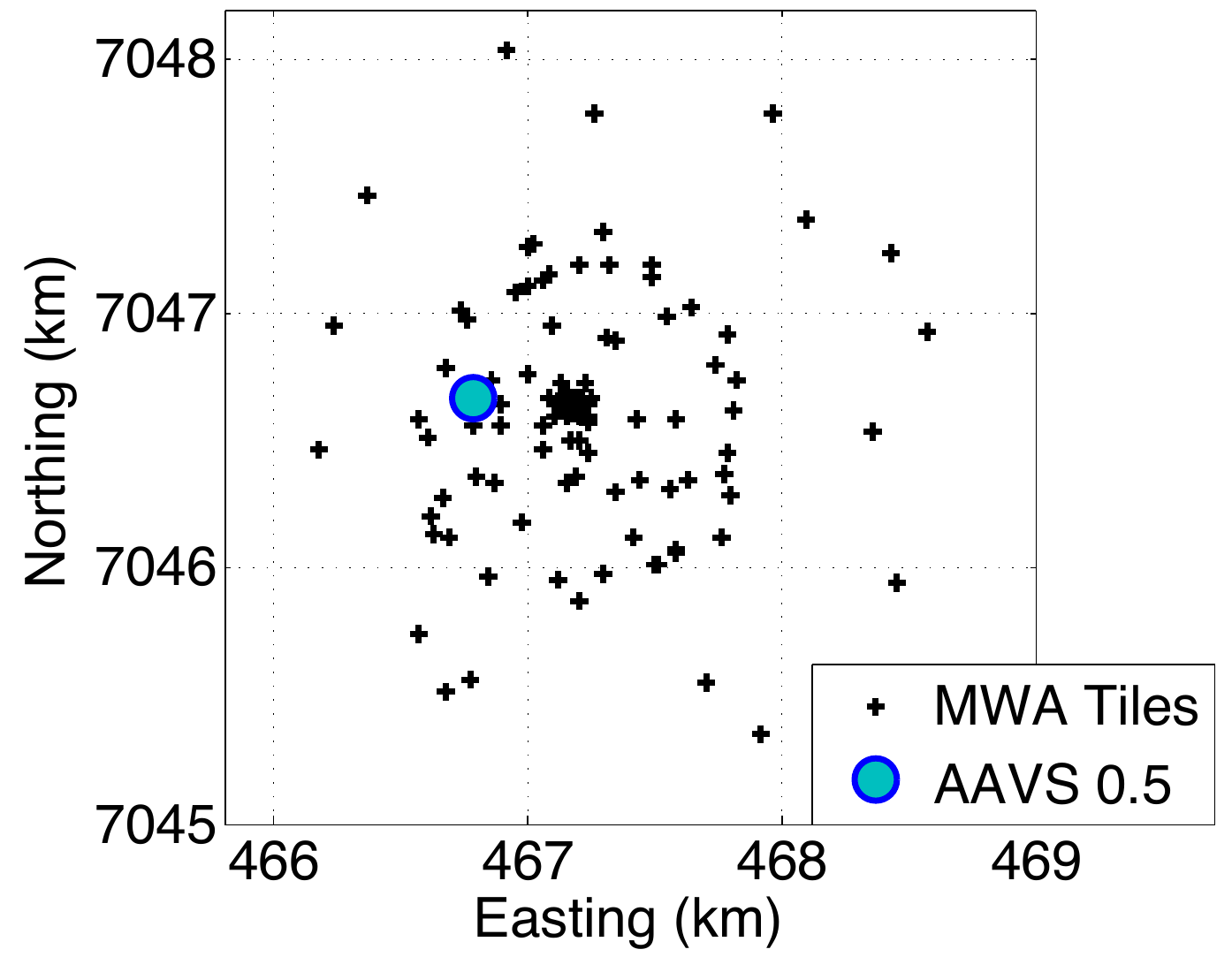}
 \caption{A map showing the AAVS\,0.5 location (blue circle) and MWA telescope tiles (black cross) in Easting/Northing for Map Grid Australia UTM zone 50. }\label{fig:AAVS05_mwa}
 \end{center}
 \end{figure}

To obtain antenna gains, amplitude-only solutions are sufficient. Note that this method is analogous to the well-known ``three-antenna'' measurement technique~\cite{Kummer_1978}. 
With $N$ elements, $N(N-1)/2$ pair-wise combinations (``baselines'') are available, leading to an overdetermined system solvable with a least-squares method. As shown in Fig.~\ref{fig:AAVS05_mwa}, the AAVS\,0.5 is co-located with the Murchison Widefield Array (MWA) radio telescope~\cite{2013PASA...30....7T, Lonsdale_2009}  to take advantage of this multi-baseline arrangement (with $N=128$) to minimize calibration uncertainties. Each of the MWA's 128 ``tiles'' is a 16-antenna array of bow-tie dipoles.

By selecting interferometry baselines larger than a certain minimum, the dominant spatially-extended Galactic noise is no longer correlated and  ``bright'' point-like sources are detectable.
This may be illustrated by examining the spatial coherence function due to a distributed source (Ch.~1 in \cite{taylor_synthesis_1999}):
\begin{eqnarray}
V_{12}=\int\int I (l,m) e^{-j2\pi(u_{12}l+v_{12}m)} dl dm,
\label{eqn:visibility2D}
\end{eqnarray}
where $V_{12}$ is the complex visibility on the baseline between antennas 1 and 2, $I (l,m)$ is the sky brightness distribution, $u,v$ are the baseline vector components normalized to wavelength and $l,m$ are the direction cosines with respect to $\hat{u},\hat{v}$ of a unit vector $\hat{s}=l\hat{u}+m\hat{v}+\sqrt{1-l^2-m^2}\hat{w}$ where $\hat{w}$ is a unit vector normal to the $uv$ plane (i.e., points to the ``phase center'').

Simplifying further to the 1-D case, (\ref{eqn:visibility2D}) becomes
\begin{eqnarray}
V_{12}=\int I (l) e^{-j2\pi(u_{12}l)} dl,
\label{eqn:visibility1D}
\end{eqnarray}
where $l=\sin\theta'$ and $\theta'$ is the angle between $\hat{w}$ and $\hat{s}$. We recognize $V_{12}$ and $I (l)$ as a Fourier transform pair. Hence, structures with large $l$ extent are mostly visible for  small~$u_{12}$. 
From MWA observing experience, with its small-diameter (3.3\,m) tiles, correlated Galactic noise is measurable on baselines of length~$<30\lambda$ and should therefore be excluded.\footnote{Some readers will recognize a similar situation in a highly multipath environment: the signal envelope is uncorrelated at locations separated in distance by $>\sim1\lambda$ \cite{Rap01}.}

It can be shown (see Ch.~9 of \cite{taylor_synthesis_1999}) that after calibration of the measured cross-correlation products, the standard deviation in flux density for each baseline is inversely proportional to the square root of the products of the $A/T$ for the AUTs in question:
\begin{equation}
\Delta S_{ij}\approx\sqrt{\frac{2}{k^2(A/T)_{i}(A/T)_{j}Bt_{\rm acc}}},
\label{eqn:DeltaS}
\end{equation}
where $k$ is Boltzmann's constant, $B$ is the bandwidth and $t_{\rm acc}$ is the accumulation time.
Thus, by measuring $\Delta S$ for all baselines for which the Galactic noise is uncorrelated, $A/T$ for each AUT may be solved. 

\section{Characterization Results}
\label{sec:results}
\begin{figure}[h]
\begin{center}
\includegraphics[width=2.1in]{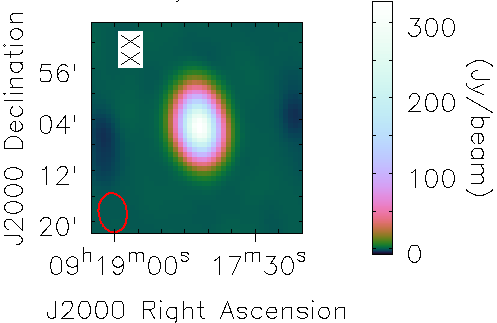}
\caption{X-polarization cross-correlation image of Hydra~A at 119\,MHz, using only baselines between MWA tiles and the AAVS\,0.5 array. The red ellipse (bottom-left corner) shows the size of the cross-correlation beam.} 
 \label{fig:AAVS_image}
\end{center}
\end{figure}

To measure $A/T $ through cross-correlation, we point the interferometer towards Hydra~A, a compact calibrator source that is sufficiently bright for a good S/N calibration solution to be obtained from a 2 minute snapshot observation (\prettyref{fig:AAVS_image}). The AAVS\,0.5 and MWA tile pointing is achieved with 5-bit, time-delay analog RF beamformers. The data is channelised prior to correlation into channels of width $B$=40\,kHz, and averaged post-correlation to $t_{\rm acc}$=4\,s. For each channel, $\Delta S_{ij}$ is calculated for all samples in the snapshot period.

\subsection{Sensitivity}
\label{sub:sensitivity}
\begin{figure}[h]
\begin{center}
\noindent
  \includegraphics[width=0.95\columnwidth]{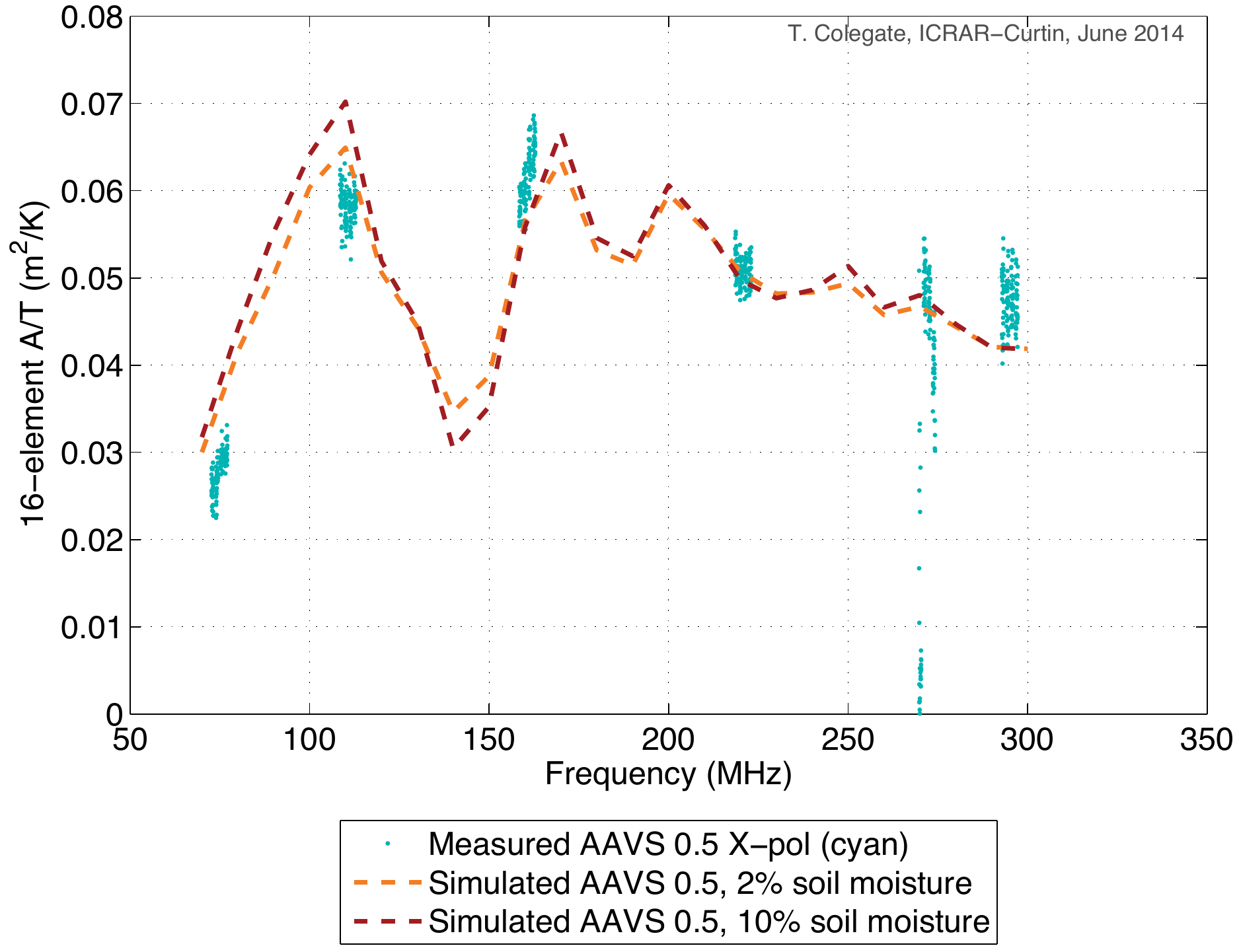}
    \includegraphics[width=0.95\columnwidth]{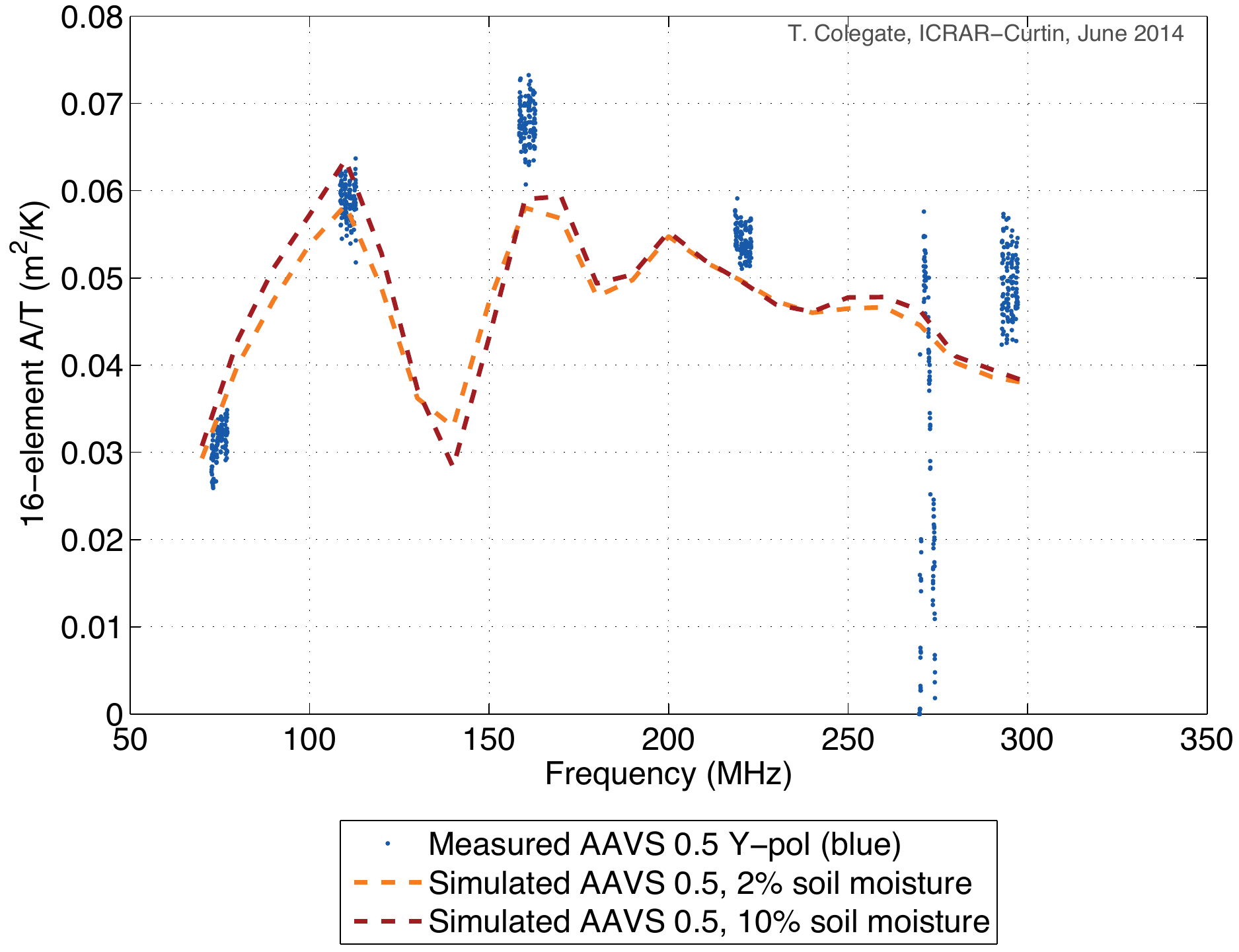}
\caption{Measured $A/T$ for a 2 minute observation of Hydra~A starting at 22-May-2014 17:30:32 at the MRO, X-polarization (East--West arm) top, Y-polarization (North--South arm) bottom. Each data point is a 40\,kHz channel measurement.
The AAVS\,0.5 array pointing is Az=0\degree, El=75.4\degree.}
 \label{fig:AonT-vs-freq}
\end{center}
\end{figure}

\prettyref{fig:AonT-vs-freq} shows an $A/T$ measurement for a snapshot, made on 22 May 2014, where the channels comprising the 30.72\,MHz bandwidth have been grouped at 6 points across the available 75--300\,MHz frequency range.
The top panel shows the X (E--W) antenna polarization in cyan dots, the lower panel shows the Y (N--S) polarization in blue dots.
In this observation, both the AAVS\,0.5 array and the MWA tiles were pointed at Hydra~A, at azimuth (clockwise from North) Az=0\degree, elevation El=75.4\degree{}. 
The scattering of data points at $\sim$270\,MHz is due to the persistent satellite-based RFI corrupting the measurement in these channels.

\prettyref{fig:AonT-vs-freq} also shows simulated results of this measurement as dashed curves; these follow \prettyref{eqn:AonT-simple} and include measured receiver temperature, and the sky model~\cite{2008MNRAS.388..247D} and array gain pattern at the time of observation. 
We used full-wave electromagnetic (FEKO\footnote[4]{http://www.feko.info/}) simulation to determine the array gain pattern for each polarization, and with a ground of either 2\% or 10\% moisture to represent a reasonable range of the likely soil moisture level experienced at the MRO~\cite{SutHal14}. 

The measured and simulated results show good agreement at all frequency points. With the exception of the corrupted data at $\sim$270\,MHz, the tight clustering of the results at each frequency point gives us confidence in the measurement process.

\subsection{Beam pattern}
\begin{figure}[h]
\begin{center}
\noindent
  \includegraphics[width=0.95\columnwidth]{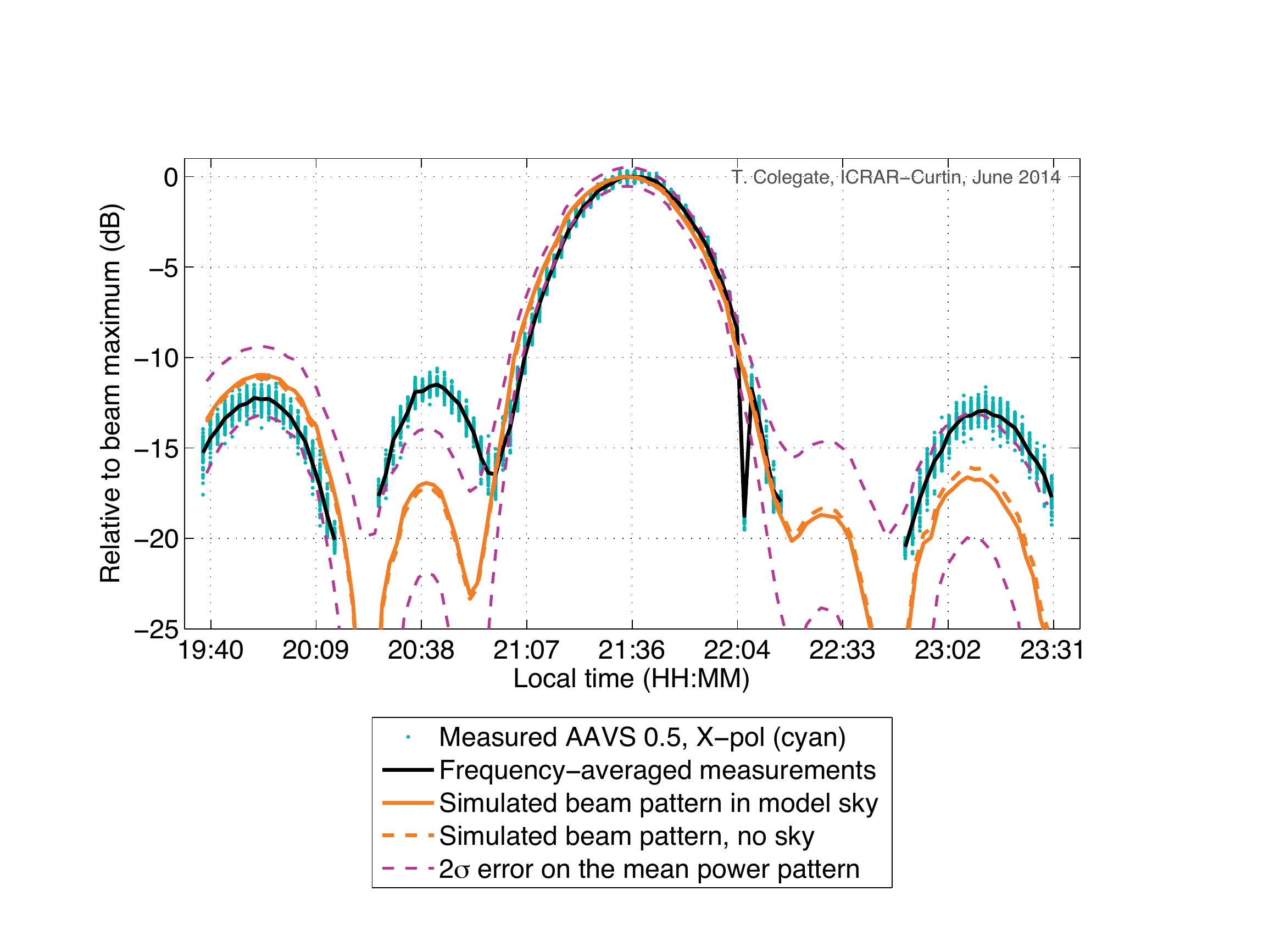}
\caption{The AAVS\,0.5 X-polarization beam pattern at 220\,MHz for Az=0\degree, El=75.4\degree{} pointing. Hydra~A  was continuously observed on 21-Mar-2014 with 2 minute snapshots and 5.12\,MHz bandwidth. Each data point is a 40\,kHz channel measurement. The black curve is the same data, frequency-averaged for each snapshot. The oranges curves are simulated (``error-free'') beam patterns for 2\% moisture in the presence of model sky (solid) or no sky (dashed). The purple curve is the $2\sigma$ uncertainty on the beam pattern.}
 \label{fig:pattern}
\end{center}
\end{figure}

Sensitivity can also be measured at Az,~El angles away from the beam pointing direction enabling investigation of beam pattern. This is achieved by keeping the AAVS\,0.5 beam pointed in a fixed direction while tracking the calibrator source with the MWA. 
\prettyref{fig:pattern} shows the X-polarization beam for such a measurement, where the MWA tracks Hydra~A over a 4 hour period and the AAVS\,0.5 beam remains pointed at Az=0\degree, El=75.4\degree. 
For each snapshot and channel, calibration is performed and $A/T$ is calculated for the AAVS\,0.5 beam at the direction of Hydra~A.
The data is normalized to the beam maximum. 
For clarity, only the 2\% moisture case is shown, as the difference with the 10\% case is not significant.

\prettyref{fig:pattern} is a slice through the AAVS\,0.5 beam pattern corresponding to the Az,~El trajectory of Hydra~A; the simulated results in \prettyref{fig:pattern} are for the same trajectory. The solid orange curve simulates the real-world scenario where the sky ``seen'' by the AAVS\,0.5 beam varies  as a function of time, inducing changes in $T_{\rm sys}$. The dashed curve is simply the beam pattern (i.e. no sky is present). 

Towards the end of the observation ($\sim$23:10), the sidelobe level from the model sky curve is decreasing relative to the ``no sky'' curve; this is caused by a reduction in $A/T$ due to increasing sky noise. 
However, the small difference between the two beam simulations is encouraging, indicating that the measured beam pattern (which intrinsically varies with time) is representative of the actual beam pattern, at least for the sky near Hydra~A. 

We also estimate errors in the beam pattern introduced due to the analog RF beamforming system. Our estimate of the errors is calculated from measured phase ($\sigma_{\phi}=0.069$) and amplitude ($\sigma_{a}=0.10$)  errors. The calculation for the ensemble mean beam pattern (sidelobe level) that incorporates these errors is well-known~\cite{Zucker_ant2_1969_ch21,Mailloux_2005_ch7,Hansen_1998_ch7}; for our results there is no significant difference to the error-free beam pattern. We plot in purple the estimated $\pm2\sigma$ error on the mean pattern, where $\sigma$ is approximated using the formula in \cite{1138220} for large $N$.
  These curves show that main beam and most of the sidelobes are within $2\sigma$ of the mean beam pattern; the exception being the first sidelobe at 20:38, which is $\sim4\sigma$ from the mean beam pattern.

\section{Conclusion}
\label{ref:concl}
This work demonstrates far-field sensitivity and beam pattern characterization of a low-frequency aperture array. The measurements are in-situ, both in terms of physical location and intended astronomical use. The interferometric observation of astronomical sources using the MWA radio telescope and the AAVS\,0.5 array under test results in measurements calibrated to an absolute scale.
The sensitivity measurements and their frequency-dependent trends are consistent with simulated results. 
The measured and simulated beam patterns generally show good agreement as to the location and height of the sidelobes within the error. 
This type of characterization will be essential for verifying the performance of future prototype low-frequency aperture arrays deployed at the MRO.

\section*{ACKNOWLEDGEMENT}
We acknowledge the Wajarri Yamatji people as the traditional owners of the Murchison Radio Astronomy Observatory site. The MRO is operated by CSIRO, whose assistance we acknowledge.  The AAVS\,0.5 operates as an external instrument of the MWA telescope and we thank the MWA project and personnel for their support.

\bibliographystyle{IEEEtran}
\bibliography{IEEEabrv,ATSbib_SKA_from_AAVS05_paper}
%

\end{document}